\documentclass[twocolumn,showpacs,aps,prd,floatfix]{revtex4}
\usepackage{graphicx}
\usepackage{dcolumn}
\usepackage{epsfig}
\usepackage{psfrag}
\usepackage{amssymb, amsmath}
\usepackage{longtable}
\usepackage{bm}
\usepackage{relsize}
\usepackage{xspace}
\long\def\inst#1{\par\nobreak\kern 4pt\nobreak
    {\it #1}\par\vskip 10pt plus 3pt minus 3pt}
\def\Y#1S{\ensuremath{\Upsilon{(#1S)}}\xspace}
\def\FourS {\Y4S}
\def\TwoS {\Y2S}
\def\ThreeS {\Y3S}
\def\babar{\mbox{\slshape B\kern-0.1em{\smaller A}\kern-0.1em
    B\kern-0.1em{\smaller A\kern-0.2em R}}}
\def\pep2{PEP-II}
\def\Kp    {\ensuremath{K^+}\xspace}
\def\Km    {\ensuremath{K^-}\xspace}
\def\piz   {\ensuremath{\pi^0}\xspace}
\def\pip   {\ensuremath{\pi^+}\xspace}
\def\pim   {\ensuremath{\pi^-}\xspace}
\def\Dz{D^0}
\def\KS    {\ensuremath{K^0_{\scriptscriptstyle S}}\xspace}
\def\D       {\ensuremath{D}\xspace}
\def\Dp      {\ensuremath{D^+}\xspace}
\def\Dm      {\ensuremath{D^-}\xspace}
\def\Dz      {\ensuremath{D^0}\xspace}
\def\Dzb     {\ensuremath{\Dbar^0}\xspace}
\def\DzDzb   {\ensuremath{\Dz {\kern -0.16em \Dzb}}\xspace}
\def\DpDm    {\ensuremath{\Dp {\kern -0.16em \Dm}}\xspace}
\def\Dstar   {\ensuremath{D^*}\xspace}
\def\Dstarb  {\ensuremath{\Dbar^*}\xspace}

\def\BR         {{\ensuremath{\cal B}\xspace}}
\def\invfb   {\ensuremath{\mbox{\,fb}^{-1}}\xspace}
\def\Dbar    {\kern 0.2em\overline{\kern -0.2em D}{}\xspace}
\def\Db      {\ensuremath{\Dbar}\xspace}
\def\DDb     {\ensuremath{\D {\kern -0.16em \Db}}\xspace}
\def\DDbX     {\ensuremath{\D {\kern -0.16em \Db}X}\xspace}
\def\DDstarb     {\ensuremath{\Dstar}\ensuremath{\Dbar}\xspace}
\def\DstarDstarb     {\ensuremath{\Dstar{\kern -0.16em \Dstarb}}\xspace}

\def\D       {\ensuremath{D}\xspace}
\def\DDb     {\ensuremath{\D {\kern -0.16em \Db}}\xspace}
\def\DDstarb     {\ensuremath{\Dstar}\ensuremath{\Dbar}\xspace}
\def\DstarDstarb     {\ensuremath{\Dstar{\kern -0.16em \Dstarb}}\xspace}
\def\Ds      {\ensuremath{D^{\pm}_s}\xspace}

\def\Dsp      {\ensuremath{D^+_s}\xspace}
\def\Dsm      {\ensuremath{D^-_s}\xspace}
\def\Dssp     {\ensuremath{D^{*+}_s}\xspace}
\def\Dssm     {\ensuremath{D^{*-}_s}\xspace}
\def\Dssqp     {\ensuremath{D^{(*)+}_s}\xspace}
\def\Dssqm     {\ensuremath{D^{(*)-}_s}\xspace}
\def\DspDsm     {\ensuremath{D^+_s}\ensuremath{D^-_s}\xspace}
\def\DsspDsm     {\ensuremath{D^{*+}_s}\ensuremath{D^-_s}\xspace}
\def\DsspDssm     {\ensuremath{D^{*+}_s}\ensuremath{D^{*-}_s}\xspace}
\def\DssqpDssqm     {\ensuremath{D^{(*)+}_s}\ensuremath{D^{(*)-}_s}\xspace}
\def\Dssq      {\ensuremath{D^{(*)\pm}_s}\xspace}

\def\MM{M^2_{\rm rec}}

\newcommand{\gevc}{\ensuremath{{\mathrm{\,Ge\kern -0.1em V\!/}c}}\xspace}
\newcommand{\mevc}{\ensuremath{{\mathrm{\,Me\kern -0.1em V\!/}c}}\xspace}
\newcommand{\gevcc}{\ensuremath{{\mathrm{\,Ge\kern -0.1em V\!/}c^2}}\xspace}
\newcommand{\mevcc}{\ensuremath{{\mathrm{\,Me\kern -0.1em V\!/}c^2}}\xspace}
\newcommand{\mev}{\ensuremath{\mathrm{\,Me\kern -0.1em V\!}}\xspace}
\newcommand{\gev}{\ensuremath{\mathrm{\,Ge\kern -0.1em V\!}}\xspace}
\newcommand{\gevcccc}{\ensuremath{{\mathrm{\,Ge\kern -0.1emV^2\!/}c^4}}\xspace}

\begin{document}
\newcommand{\BaBarYear}    {10}
\newcommand{\BaBarNumber}  {015}
\newcommand{\BABARProcNumber} {\phantom{14}}
\newcommand{\SLACPubNumber} {14209}
\newcommand{\LANLNumber} {yymm.nnnn [hep-ex]}
\newcommand{\BaBarType}      {PUB}  
\begin{flushleft}
\babar-\BaBarType-\BaBarYear/\BaBarNumber \\
SLAC-PUB-\SLACPubNumber 
\end{flushleft}

\title{\boldmath Exclusive Production of \DspDsm, \DsspDsm, and \DsspDssm via  $e^+ e^-$ Annihilation with Initial-State-Radiation}
\author{P.~del~Amo~Sanchez}
\author{J.~P.~Lees}
\author{V.~Poireau}
\author{E.~Prencipe}
\author{V.~Tisserand}
\affiliation{Laboratoire d'Annecy-le-Vieux de Physique des Particules (LAPP), Universit\'e de Savoie, CNRS/IN2P3,  F-74941 Annecy-Le-Vieux, France}
\author{J.~Garra~Tico}
\author{E.~Grauges}
\affiliation{Universitat de Barcelona, Facultat de Fisica, Departament ECM, E-08028 Barcelona, Spain }
\author{M.~Martinelli$^{ab}$}
\author{A.~Palano$^{ab}$ }
\author{M.~Pappagallo$^{ab}$ }
\affiliation{INFN Sezione di Bari$^{a}$; Dipartimento di Fisica, Universit\`a di Bari$^{b}$, I-70126 Bari, Italy }
% ...
\collaboration{The \babar\ Collaboration}
\noaffiliation
\date{\today}

\begin{abstract}

We perform a study of
exclusive production of  \DspDsm, \DsspDsm, and \DsspDssm final states in initial-state-radiation events from
$e^+ e^-$ annihilations at a center-of-mass energy near 10.58 \gev, to search for 
charmonium $1^{--}$ states.
The data sample corresponds to an integrated luminosity of 525~\invfb
and was recorded by the \babar\ experiment at the \pep2 storage ring.
The \DspDsm, \DsspDsm, and \DsspDssm mass spectra show evidence of the known $\psi$ resonances.
Limits are extracted for the branching ratios of the decays $X(4260)\to \DssqpDssqm$. 
\end{abstract}


\maketitle

\section{Introduction}
The surprising discovery of new states decaying to  $J/\psi \pi^+ \pi^-$
\cite{charmonium,babar_Y} has renewed interest in the field of charmonium spectroscopy, 
since not all the new resonances are easy
to accommodate in the quark
model.
Specifically, the \babar\ experiment discovered a broad state, $X(4260)$,
decaying to $J/\psi \pi^+ \pi^-$, in the initial-state-radiation (ISR)
reaction $e^+ e^- \to \gamma_{ISR} X(4260)$. 
Its quantum numbers  $J^{PC}=1^{--}$
are inferred from the single virtual-photon production.
Enhancements in the $\psi(2S) \pi^+ \pi^-$ mass
distribution at $4.36\ \gevcc$~\cite{babar_Y2,belle_Y2} and $4.66\ \gevcc$~\cite{belle_Y2} have been observed 
for the reaction $e^+ e^- \to \gamma_{ISR} \psi(2S) \pi^+ \pi^-$. 
Charmonium
states at these masses would be expected \cite{eichten,barnes} to decay predominantly to \DDb, \DDstarb, or
\DstarDstarb. 
It is peculiar that the decay rate to the hidden charm final state
$J/\psi \pi^+ \pi^-$ is much larger for the $X(4260)$ than for the higher-mass charmonium states 
(radial excitations)~\cite{mo}.
Many theoretical interpretations for the $X(4260)$ have been proposed, 
including unconventional scenarios:  
quark-antiquark gluon hybrids~\cite{Y4260-hybrid}, baryonium~\cite{qiao}, tetraquarks~\cite{maiani},
and hadronic molecules~\cite{Y4260-molecule}. 
If the $X(4260)$ were a diquark-antidiquark state $[cs][\bar c \bar s]$, as proposed
by L. Maiani {\it et al.}~\cite{maiani}, this state would predominantly decay to \DspDsm.
For a discussion and a list of 
references see, for example, Ref.~\cite{swanson}.

In this paper, we present a study of the ISR production of \DspDsm, \DsspDsm, and \DsspDssm~\cite{conj} final 
states, and search for evidence of 
charmonium states and 
resonant structures. This follows earlier \babar \  measurements of the cross section of 
\DDb~\cite{babar_dd} and of \DDstarb and \DstarDstarb production
~\cite{babar_dstard} and studies of these final states~\cite{belle_dd,dstar2} by the Belle Collaboration. Recently the CLEO
Collaboration~\cite{cleo} studied $e^+ e^-$ annihilation to  \DspDsm, \DsspDsm, and \DsspDssm final states at
center of mass energies from threshold up to 4.3 GeV \gevcc.
In the present analysis we extend these measurements up to 6.2 \gevcc.

This paper is organized as follows. 
A short description of the \babar\ experiment is given in Section II, and data selection is described in Section III.
In Sections IV, V, and VI we present studies of the  \DspDsm, \DsspDsm, and  \DsspDssm 
final states, respectively.
Fits to the three final states are described in Section VII, and in Section VIII 
we present limits on the decay of the $X(4260)$ to \DssqpDssqm. 
A summary and conclusions are found in Section IX.

\section{The \babar\ experiment}
This analysis is based on a data sample
of 525~$\invfb$ recorded mostly at the
\FourS resonance and 40 MeV below the resonance by the \babar\ detector at the \pep2
asymmetric-energy $e^+e^-$ storage rings. The sample includes also 15.9~$\invfb$ and 31.2~$\invfb$
collected at the \TwoS and \ThreeS respectively, and 4.4~$\invfb$ above the \FourS resonances.
The \babar\ detector is
described in detail elsewhere~\cite{babar}. We mention here only the components of the 
detector that are used in the present analysis.
Charged particles are detected
and their momenta measured with a combination of a 
cylindrical drift chamber (DCH)
and a silicon vertex tracker (SVT), both operating within a
1.5 T magnetic field of a superconducting solenoid. 
Information from
a ring-imaging Cherenkov detector is combined with specific ionization measurements from the 
SVT and DCH to identify charged kaon and pion candidates. 
The efficiency for kaon identification is 90\% while the rate for a kaon being misidentified as a pion is 2\%.
Photon energies are measured with a 
CsI(Tl) electromagnetic calorimeter (EMC).

\section{Data Selection}
For each candidate event, we first reconstruct a \DspDsm pair. While one of
the \Dsp is required to decay to \Kp\Km\pip, we include three different decay
channels for the second \Dsm (see Table~\ref{tab:tab1}).
\begin{table}[tbp]
\caption{Reconstructed decay channels for the two \Ds mesons in each event.}
\label{tab:tab1}
\begin{center}
\vskip -0.2cm
\begin{tabular}{lll}
\hline
\hline \noalign{\vskip1pt}
Channel & First $D_s$ decay   & Second $D_s$ decay \cr
\hline \noalign{\vskip2pt}
(1) & \Kp \Km \pip & \Kp \Km \pim  \cr
(2) & \Kp \Km \pip & \Kp \Km \pim \piz  \cr
(3) & \Kp \Km \pip & \KS \Km  \cr
\hline
\hline
\end{tabular}
\end{center}
\end{table}
\Dssp decays are reconstructed via their decay $\Dssp \to \Dsp \gamma$.

For all final states,
events are retained if the number of well-measured charged tracks having
a transverse momentum greater than 0.1 \gevc is exactly
equal to the total number of charged daughter particles. 
EMC clusters with a minimum energy of 30 \mev that are not associated with a charged track
are identified as photons.
Candidates for the decay $\piz \to \gamma \gamma$ are kinematically constrained 
        to the \piz mass. For $\KS \to \pip\pim$ candidates we apply vertex and mass
constraints.
The tracks corresponding to the charged daughters of each \Dsp candidate are constrained to come from
a common vertex.  Reconstructed \Dsp candidates with a fit 
probability greater than 0.1\% are retained.
Each $\DspDsm$ pair is refit to a common vertex with
the constraint that the pair originates from the $e^+ e^-$ interaction region.
Only candidates with a $\chi^2$ fit probability greater than 0.1\% are retained.
For each event we consider all combinations.

ISR Monte Carlo (MC)~\cite{isr} events for each final state are fully simulated using the
GEANT4 detector simulation package~\cite{geant}, and they are processed through the same
reconstruction and analysis chain as the data. 

We select \Dsp and \Dssp candidates using the reconstructed \Dsp mass 
and the mass difference, which for $\Dsp \to \Kp \Km \pip$ is defined as
$\Delta m(\Dsp \gamma) \equiv m(\Kp \Km \pip \gamma) - m(\Kp \Km \pip)$. The \Dsp
parameters are obtained by fitting the relevant mass spectra using a polynomial 
for the background and a single Gaussian for the signal. 
For \Dssp, we use the PDG~\cite{pdg} mass and a Gaussian width $\sigma$ = 6 \mevcc obtained by MC simulations.
Events are selected within $\pm 2.0 \sigma$ from the 
fitted central values.
The $D^{(*)+}_s$ candidate three-momentum is determined from the summed 
three-momenta of its decay particles. The nominal $D^{(*)+}_s$ mass~\cite{pdg} is used to 
compute the energy component of its four-momentum.

The ISR photon is preferentially emitted at small angles 
with respect to the beam axis, and it escapes detection in the majority of ISR
events. Consequently, the ISR photon is treated as a missing particle. 
We define the squared mass $\MM$ recoiling against the \DspDsm, \DsspDsm, and \DsspDssm 
systems using the four-momenta of the beam
particles $p_{e^\pm}$ and of the reconstructed $D^{(*)\pm}_s$ $p_{D^{(*)\pm}_s}$:
\begin{equation}
\MM \equiv ( p_{e^-}+p_{e^+}-p_{\Dssqp}-p_{\Dssqm})^2.
\end{equation}
This quantity 
should peak near zero for both ISR events and for exclusive production of 
$e^+ e^- \to \DssqpDssqm$. For exclusive production,
the \DssqpDssqm  mass distribution peaks at the kinematic limit. 
We reject exclusive
events by requiring the \DssqpDssqm mass to be below 6.2  \gevcc and select
ISR candidates by requiring $| \MM |<0.8 \ \gevcccc$.

\begin{figure*}[!htb]
\begin{center}
\includegraphics[width=14cm]{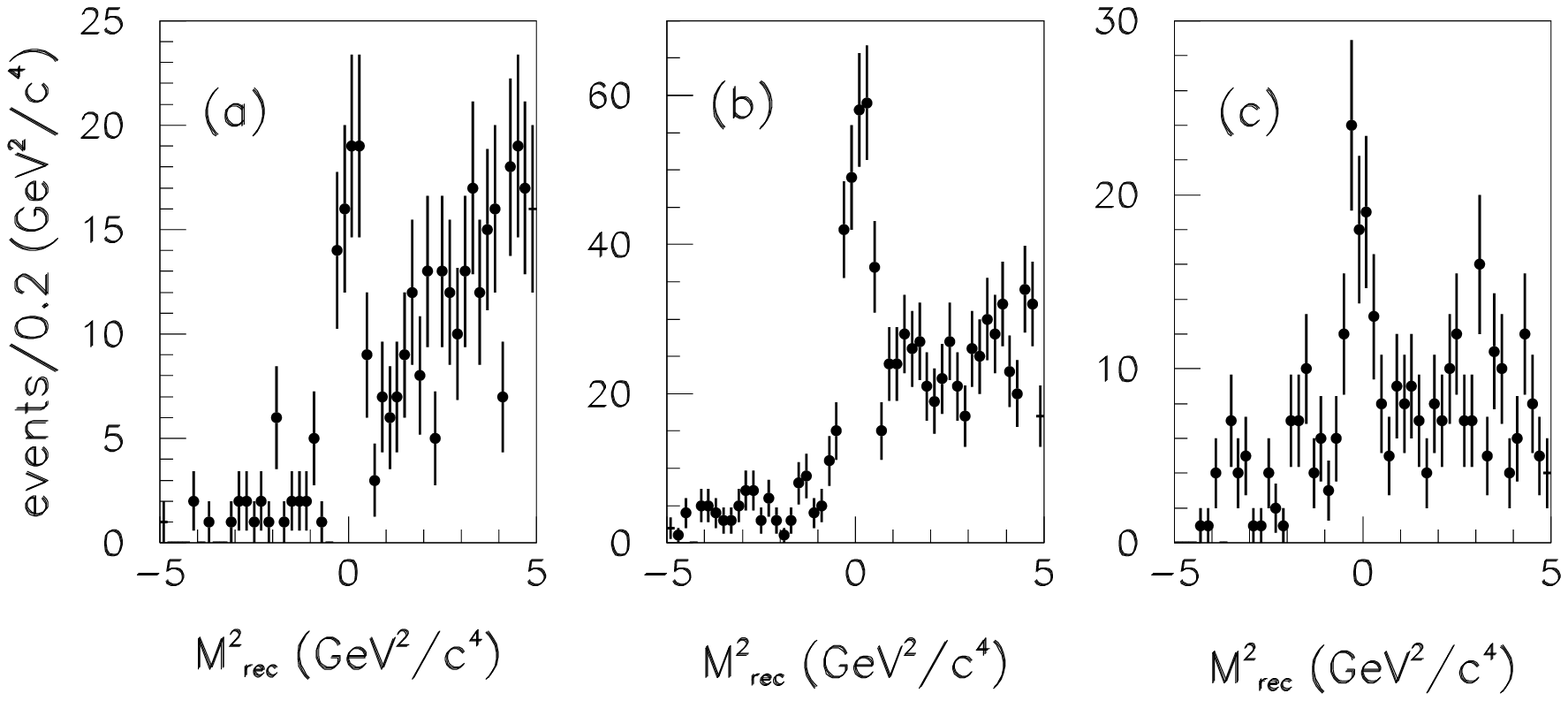}
\caption{Distributions of $\MM$ for the (a) \DspDsm, (b) \DsspDsm, and (c) \DsspDssm final states.}
\label{fig:fig1}
\end{center}
\end{figure*}
We allow additional \piz and photon candidates due to radiative 
or background  photons.  This introduces multiple candidates
in the reconstruction of the different channels.
For channel (1)-(2) ambiguities, each [\Kp \Km \pip (\piz)] combination
is considered as a candidate for both channels.
For all channels, each [\Dsp, \Dsm, ($\gamma$), ($\gamma$)] 
combination is considered as a candidate for 
\DspDsm, \DsspDsm, and \DsspDssm.

To discriminate among the different \Dsp channels and \DssqpDssqm final states, and to separate signal from background, we make use of a likelihood ratio test:
\begin{equation}
L = \sum_{i=1}^N \log(PDF^S_i) -\sum_{i=1}^N \log(PDF^B_i) 
\end{equation}
\noindent
where $N$ is the number of discriminating variables, while $PDF^S_i$ and $PDF^B_i$ are normalized distributions 
describing signal and background, respectively.
Signal $PDF^S_i$ are obtained from MC simulations. Background $PDF^B_i$ are obtained from the data. Since the ISR signal is very 
small compared to the entire data set of candidates ($<$ 0.1 \%), we use the data as the background model by relaxing all the selection criteria,
except $m(\DssqpDssqm)<6.2$ \gevcc. 

The discriminating variables used in the likelihood ratio are the following.
\begin{itemize}
\item{} The number of additional \piz candidates in the event. For decay channel (2) this number is computed after removing
the \piz from \Dsp decay. This distribution is expected to peak at zero for signal events.
\item{} The residual energy in the calorimeter, which is computed after removing any
ISR photon candidate, identified by a center-of-mass energy greater than 2.0 \gevcc. For \DsspDsm and \DsspDssm
final states, the $\gamma$ from \Dssp decays is excluded from the residual energy calculation.
This distribution is expected to peak at zero for signal events.
\item{} The distribution of $\cos \theta^*$, 
where $\theta^*$ is the polar angle of the \DssqpDssqm
system in the center-of-mass frame which peaks at $\pm 1$ for ISR events.
\item{} The momentum distribution of the \piz from the \Dsp for decay channel (2).
\item{} The $\gamma$ energy distribution from \Dssp for \DsspDsm and \DsspDssm final states.
\end{itemize}

For each \Dsp decay channel (1)-(3) and for each \DspDsm, \DsspDsm, and \DsspDssm final state, we produce a likelihood ratio test
according to Eq.~(2) and apply empirically determined cuts on $L$ in order to reduce the background and minimize the signal loss.

\section{Study of the \DspDsm final state}

MC studies demonstrate that the main background to the \DspDsm 
final state is from \DsspDsm events, which
have a larger cross section.
Therefore, we eliminate \DspDsm candidates if they are also identified as \DsspDsm candidates.
MC simulations show that  
this veto rejects about 14\% of the true \DspDsm final states and that the residual background is consistent with
\DsspDsm feedthrough.

Figure~\ref{fig:fig1}(a) shows the $\MM$  
distribution for the selected \DspDsm candidates, summed over the
the \Dsp decay channels (1)-(3).
The peak centered at zero is evidence for the ISR process.
To determine the number of signal and background events, we perform a $\chi^2$ fit to the
$\MM$ distribution. The background is approximated by a $2^{nd}$ order
polynomial. The signal lineshape is taken from \DspDsm MC simulations. 
The resulting yield
and the fitted purity $P$, defined as $P=N_{\rm signal}/(N_{\rm signal}+N_{\rm background})$ 
are summarized in Table~\ref{tab:tab2}.

\begin{table}[tbp]
\caption{Number of signal ISR candidates and purities for the different final states calculated in the range $| \MM| <0.8\ \gevcccc$.}
\label{tab:tab2}
\begin{center}
\vskip -0.2cm
\begin{tabular}{lcc}
\hline
\hline \noalign{\vskip2pt}
Final state \ & \ Signal+Background \ & \ Purity(\%) \cr
\hline \noalign{\vskip2pt}
$\DspDsm$ & 81  & 65.4 $\pm$ 5.3 \cr
\hline \noalign{\vskip2pt}
$\DsspDsm$ & 286 & 67.1 $\pm$ 2.8 \cr
\hline \noalign{\vskip2pt}
 $\DsspDssm$ & 105 & 54.3 $\pm$ 4.9 \cr
\hline
\hline
\end{tabular}
\end{center}
\end{table}

\begin{figure}[!htb]
\begin{center}
\includegraphics[width=8cm]{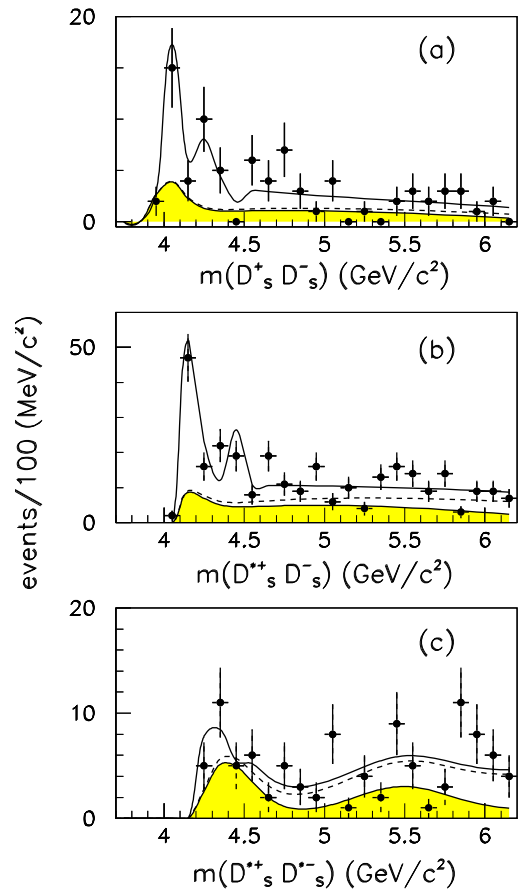}
\caption{The observed (a) \DspDsm, (b) \DsspDsm, and (c) \DsspDssm mass spectra. 
The shaded areas show the background derived from fits to the
$\MM$ sidebands. The dashed lines indicate the sum of this background and the
coherent background. The solid lines are the results from the fit 
described in Section VII.}
\label{fig:fig2}
\end{center}
\end{figure}

The \DspDsm mass spectrum, presented in Fig.~\ref{fig:fig2}(a), 
shows a threshold enhancement at the position of
the $\psi(4040)$ and a small enhancement around 4.26 \gevcc. 

We make use of the Gaussian functions to describe the presence of
the peaking backgrounds.
The \DspDsm background, taken from $\MM$ sideband events ($1.5 < |\MM| < 3.5\ \gevcccc$), is fitted to a sum of a
Gaussian function and a $3^{rd}$ order polynomial.
The fitted \DspDsm mass spectrum for these events, normalized to the 
background estimated from the fit to the $\MM$ distribution, is presented as the shaded distribution 
in Fig.~\ref{fig:fig2}(a). 

The \DspDsm reconstruction efficiency and the mass resolution for each channel have been studied
in the mass region between 4.25 and $6.25\ \gevcc$.
The \DspDsm mass resolution
is similar for decay channels (1) and (3) and slightly worse for decay channel (2) (by $\approx$ 1 \mevcc). 
It increases with \DspDsm mass from 3.5 to 5.5 MeV$/c^2$ in the mass region of the $\psi$ resonances ($<$ 5 \gevcc). 
The mass-dependent reconstruction efficiency for the \DspDsm decay channel 
$i$ ($i=1,3$), $\epsilon_i(m_{\DspDsm})$, evaluated at five different mass values,
is parameterized in terms of a $2^{nd}$ order
polynomial, and scaled to 
account for the product branching fractions for each channel, $\BR_i$~\cite{pdg}, given in
Table~\ref{tab:tab1}, 
\begin{equation}
\epsilon_i^{\cal{B}}(m_{\DspDsm}) = \epsilon_i(m_{\DspDsm}) \times \BR_i.
\end{equation}

\begin{figure*}[!htb]
\begin{center}
\includegraphics[width=14cm]{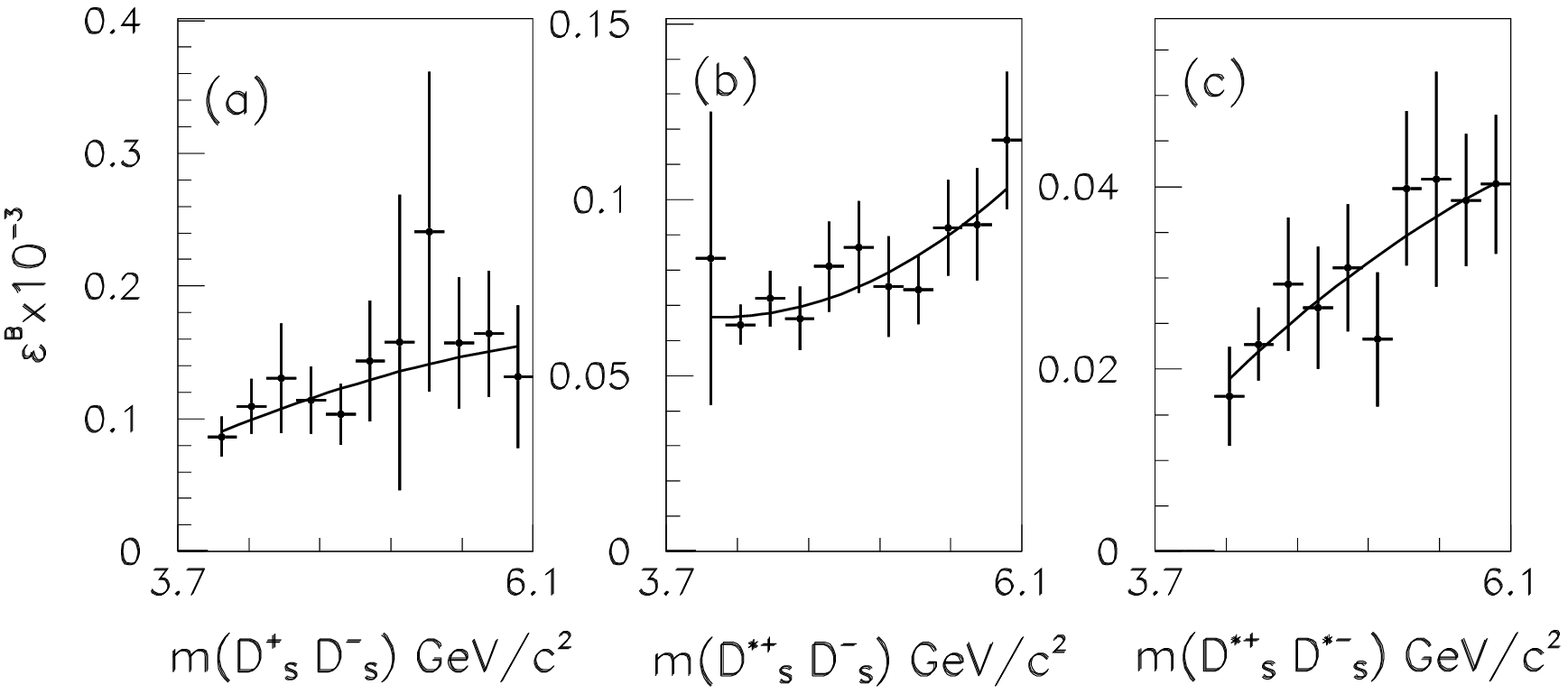}
\caption{Weighted efficiencies $\epsilon^{\cal{B}}$ for (a) \DspDsm, (b) \DsspDsm, and (c) \DsspDssm.} 
\label{fig:fig3}
\end{center}
\end{figure*}

These values are weighted by $N_i(m_{\DspDsm})$, the number of $\DspDsm$ candidates in decay channel $i$,
to compute the average efficiency as a function of $m_{\DspDsm}$,

\begin{equation}
\epsilon^{\cal{B}}(m_{\DspDsm}) = \frac{\sum_{i=1}^{3} N_i(m_{\DspDsm})}{\sum_{i=1}^{3} \frac{N_i(m_{\DspDsm})}{\epsilon^{\cal{B}}_i(m_{\DspDsm})}}.
\end{equation}
The $\epsilon^{\cal{B}}$ function for \DspDsm is shown 
in Fig.~\ref{fig:fig3}(a).
The three \DspDsm decay channels, after correcting for efficiency and branching fractions,
have yields that are consistent within statistical errors.

\begin{figure}[!htb]
\begin{center}
\includegraphics[width=8cm]{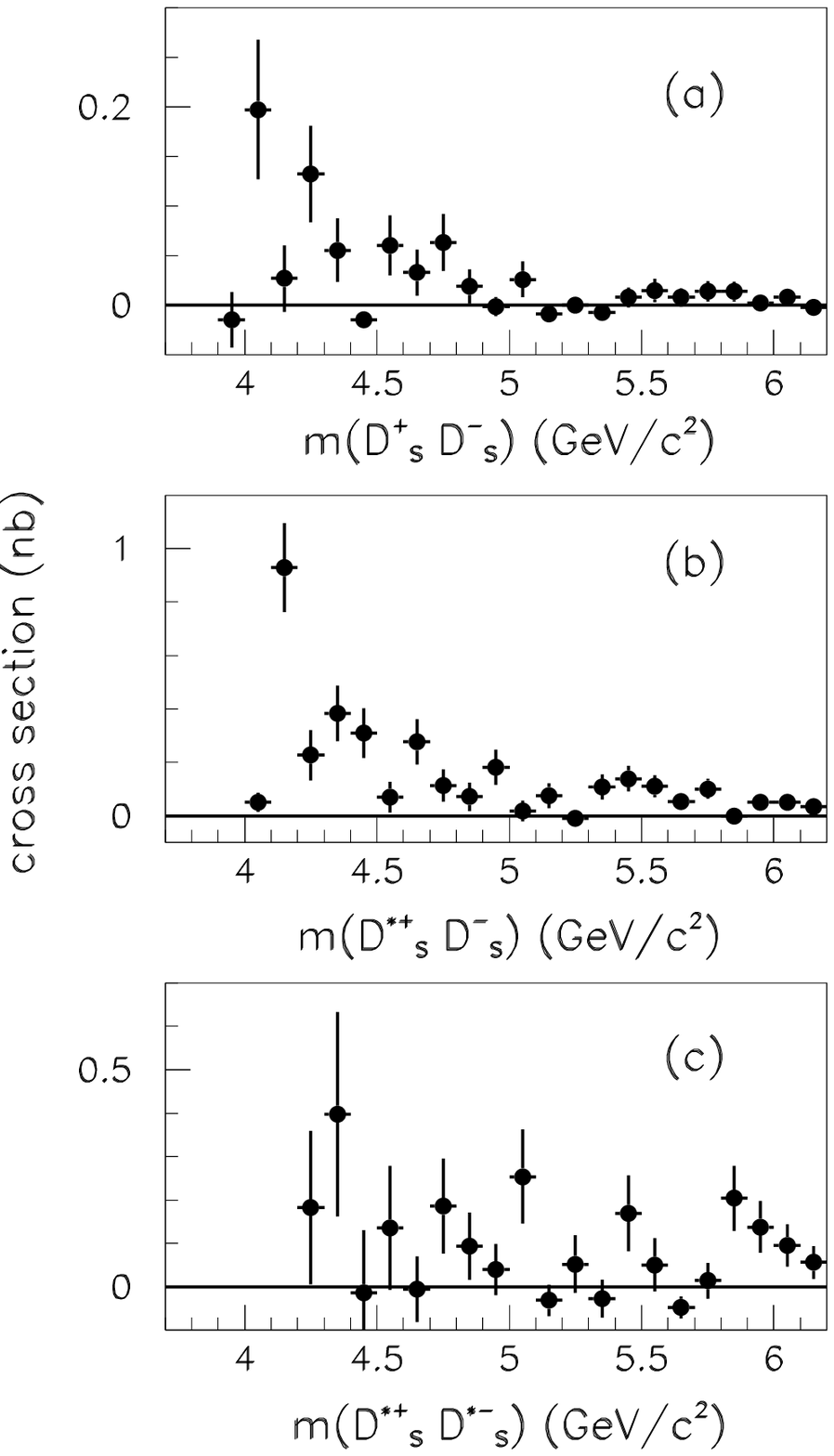}
\caption{Cross section for $e^+ e^- \to$ (a) \DspDsm, (b) \DsspDsm, and (c) \DsspDssm. The error bars correspond
to statistical errors only.} 
\label{fig:fig4}
\end{center}
\end{figure}
The $\DspDsm$ cross section is computed using
\begin{equation}
\sigma_{e^+e^-\to \DspDsm}(m_{\DspDsm}) = \frac{dN/dm_{\DspDsm}}{\epsilon^{\cal{B}}(m_{\DspDsm})
  d{\cal{L}}/dm_{\DspDsm}}, 
\end{equation}
where $dN/dm_{\DspDsm}$ is the background-subtracted yield.
The differential luminosity is computed as~\cite{benayoun}
\begin{equation}
\frac{d{\cal{L}}}{dm_{\DspDsm}} = {\cal{L}} \frac{2m_{\DspDsm}}{s} \frac{\alpha}{\pi
  x}(\ln(s/m_e^2)-1)(2-2x+x^2), 
\end{equation}
where $s$ is the square of the 
$e^+ e^-$ center-of-mass energy, $\alpha$ is the fine-structure constant, $x = 1 - m_{\DspDsm}^2/s$,
$m_e$ is the electron mass, and $\cal{L}$ is the
integrated luminosity of 525~\invfb.
The cross section for \DspDsm is shown
in Fig.~\ref{fig:fig4}(a). 
This result can be compared with QCD calculations~ref.~\cite{brodsky},  which predict  a
vanishing cross section near 5 \gevcc.

The list of systematic uncertainties for the \DspDsm cross section is summarized in Table~\ref{tab:tab3} and 
it is evaluated to be 23\%. It
includes contributions from particle identification, tracking, photon and \piz reconstruction efficiencies,
background estimates, branching fractions, and the criteria to select the
final state.
All contributions are added in quadrature. 
The  \DspDsm  systematic error is dominated by the uncertainty in the veto of the \DsspDsm
events.
\begin{table}[tbp]
\caption{Systematic uncertainties (in \%) for the evaluation of the 
\DspDsm, \DsspDsm, and  \DsspDssm cross sections.}
\label{tab:tab3}
\begin{center}
\vskip -0.2cm
\begin{tabular}{lccc}
\hline
\hline
Source &  \DspDsm & \DsspDsm & \DsspDssm \cr
\hline
Background subtraction & 18.0 & 4.2 & 4.9 \cr 
Branching fractions &  10.0 & 10.0 & 10.0 \cr
Particle identification & 5.0 & 5.0 & 5.0 \cr
Tracking efficiency & 1.4 & 1.4 & 1.4 \cr
$\pi^0$'s  and $\gamma$ & 1.1 & 2.9 & 4.7 \cr
Likelihood selection & 8.7 & 4.0 & \cr
\hline
Total & 23 & 13 & 13\cr
\hline
\hline
\end{tabular}
\end{center}
\end{table}

\section{Study of the \DsspDsm final state}
\begin{figure}[!htb]
\begin{center}
\includegraphics[width=10cm]{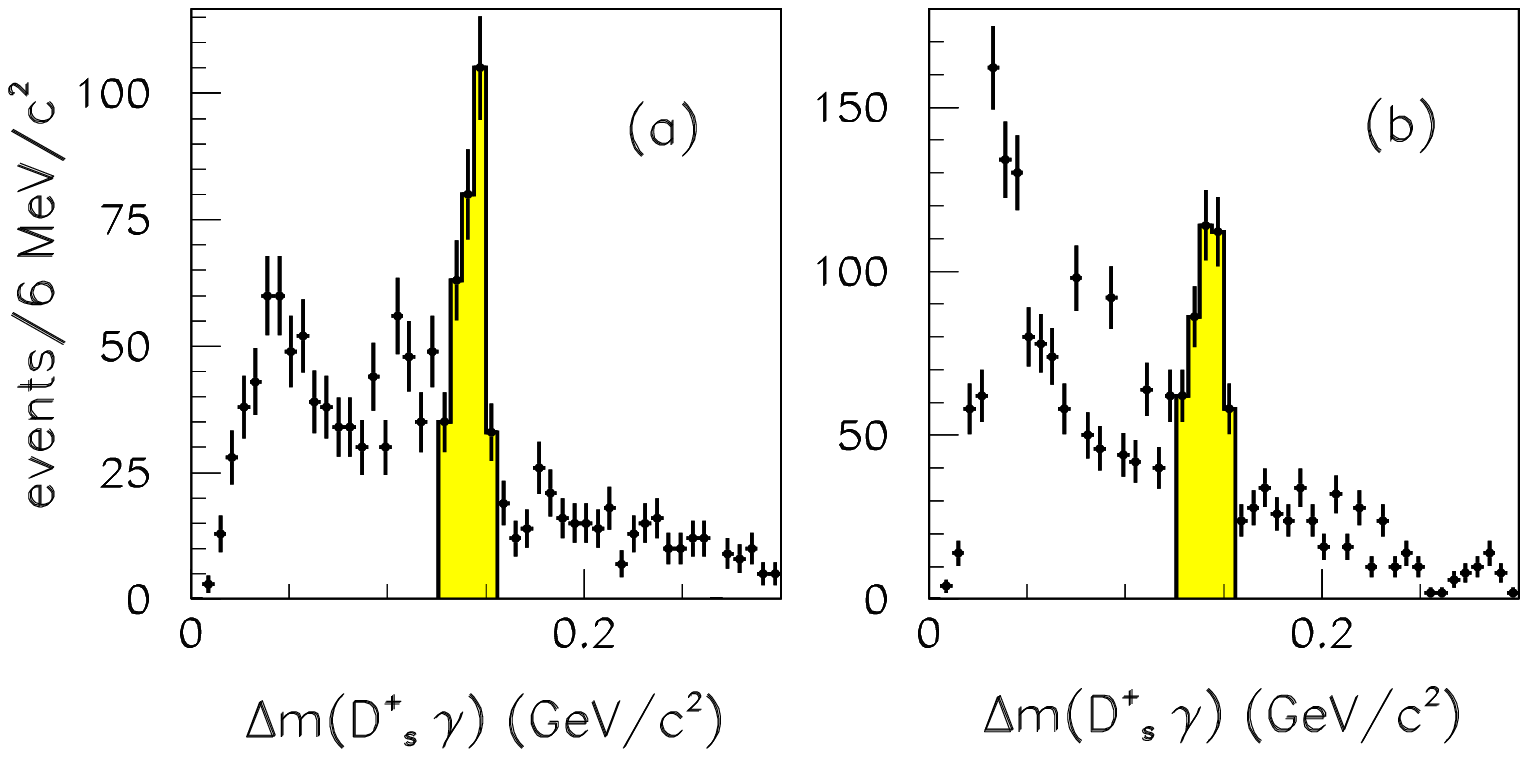}
\caption{$\Delta m$ distributions for \Dssp candidates after applying the $|\MM |<0.8\ \gevcccc$ and $m(\DssqpDssqm)<6.2$ \gevcc selections,
for the (a) \DsspDsm, and (b) \DsspDssm samples. 
The shaded regions indicate the ranges used to select
the $D^*_s$ candidates.}
\label{fig:fig5}
\end{center}
\end{figure}

A similar analysis is carried out for the \DsspDsm. 
Figure~\ref{fig:fig5}(a) shows 
the $\Delta m(\Dsp \gamma)$ 
distributions for \DsspDsm candidates passing the ISR requirements described in Sect. III. We also require the presence of a reconstructed \Dsm.

The $\MM$ distribution for  \DsspDsm candidates is shown in
Fig.~\ref{fig:fig1}(b) where a clear signal of ISR production is observed.
The number of ISR candidates and sample purity are
summarized in Table~\ref{tab:tab2}.

The \DsspDsm mass spectrum and background are shown in Fig.~\ref{fig:fig2}(b) 
and is dominated by the $\psi(4160)$ resonance. 
The \DsspDsm mass resolution
is similar for the three decay channels 
and increases with \DsspDsm mass from 7 to 8 MeV$/c^2$ in the 
mass region of the $\psi$ resonances.
The weighted efficiency $\epsilon^{\cal{B}}$ is shown in Fig.~\ref{fig:fig3}(b).
The \DsspDsm cross section is calculated using 
the method described in Sec. IV for \DspDsm. 
The result is shown in Fig.~\ref{fig:fig4}(b).
The overall systematic error for
the cross section is 13\% and is dominated by the uncertainties in the branching fractions~\cite{pdg} 
(see Table~\ref{tab:tab3}).

\section{Study of the \DsspDssm final state}

For the selection of \DsspDssm candidates, we do not make use of the likelihood 
test described in the previous sections
because no improvement for the signal to background ratio is obtained. 
Instead, we require the two photon invariant mass $m(\gamma \gamma)$
to lie outside 
the $\pi^0$ window.
 
The ambiguity in the $\gamma$'s assignment to \Dssp or \Dssm 
is resolved by choosing the \Dsp $\gamma$-\Dsm $\gamma$ combinations with
both \Dssq masses closest the expected value. Fig.~\ref{fig:fig1}(c) shows the resulting
$\MM$ distribution which shows 
clear evidence for the signal final state produced in interactions with ISR. 
For the selected the ISR signal candidates,  
we show the $\Delta m(\Dsp \gamma)$ distribution (two combinations per event) in Fig.~\ref{fig:fig5}(b).
The resulting event yield and purity are summarized in Table~\ref{tab:tab2}. 

The \DsspDssm mass spectrum and background are shown in Fig.~\ref{fig:fig2}(c). 
Due to the presence of structures, the background in this case is fitted using a $3^{rd}$ order polynomial 
and two Gaussians. 
Monte Carlo studies indicate that an important part
of this background is due to the \DsspDsm final state plus a random background $\gamma$. 
The \DsspDssm mass resolution
is similar for the three decay channels. It
increases with \DsspDssm mass from 9 to 11 MeV$/c^2$ in the mass 
region of the $\psi$ resonances.
The weighted efficiency $\epsilon^{\cal{B}}$ is shown in Fig.~\ref{fig:fig3}(c).
The \DsspDssm cross section shown in Fig.~\ref{fig:fig4}(c)
is calculated using the same method used to compute the \DspDsm cross section. 
The overall uncertainty for
the cross section is 13\% and is dominated by the uncertainties on the \Dssq branching bractions (see Table~\ref{tab:tab3}).

The \DspDsm, \DsspDsm, and \DsspDssm cross sections, where they overlap, are in good agreement with
CLEO~\cite{cleo} measurements.

\section{Fit to the Mass Spectra}
Unbinned maximum likelihood fits are performed separately to the \DspDsm, \DsspDsm, and \DsspDssm mass spectra. The likelihood function used is 
\begin{eqnarray}
L = f  \epsilon^{\cal{B}}(m) |  P(m) + c_1 W_1(m) e^{i \phi_1} + ... + c_n W_n(m)e^{i \phi_n}|^2 \nonumber\\
+ B(m)(1 - f),
\end{eqnarray}
\noindent
where $m$ is the \DssqpDssqm mass, $c_i$ and
$\phi_i$ are free parameters, $W_i(m)$ are P-wave relativistic Breit-Wigner
distributions~\cite{pdg}, $P(m)$ represents the nonresonant contribution, $B(m)$ 
is the background described in Sect. IV, $\epsilon^{\cal{B}}(m)$ is the weighted efficiency, and $f$ is the signal fraction
fixed to the values obtained fitting the $\MM$ distributions. 
In this way we allow interference between the resonances and the
nonresonant contribution $P(m)$.
The shape of the nonresonant contribution $P(m)$ is unknown; we therefore parametrize it in a simple way
as
\begin{equation}
P(m) =C(m)(a + b m),
\end{equation}
where $C(m)$ is the phase space function for \DssqpDssqm, and $a$ and $b$ are free parameters.
The size of the nonresonant production is determined by the fit.

The mass and width of the $\psi(4040)$,
$\psi(4160)$, $\psi(4415)$ and $X(4260)$ are fixed to the values reported in~\cite{pdg}. 
Resolution effects can be ignored since the widths of the resonances are
much larger than the experimental resolution. 

The three \DspDsm, \DsspDsm, and \DsspDssm likelihood functions are computed
with different thresholds, efficiencies, purities, backgrounds, and numbers of contributing resonances appropriate for
each final state. 
The results of this fits are compared to the data in Fig.~\ref{fig:fig2}, both the total 
fitted yield as well as the coherent nonresonant contribution, 
$|P(m)|^2$, ignoring any interference effects.

The fraction for each resonant contribution $i$
is defined by the following expression:
\begin{equation}
f_i = \frac {|c_i|^2 \int |W_i(m)|^2 dm}
{\sum_{j,k} c_j c_k^* \int W_j(m) W_k^*(m) dm}.
\end{equation}
The fractions $f_i$ do not necessarily add up to 1 because of interference
between amplitudes. The error for each fraction has been evaluated by propagating the 
full covariance matrix obtained by the fit. The resulting fit fractions 
are given in Table~\ref{tab:tab4}.
\begin{table}[tbp]
\caption{\DspDsm, \DsspDsm, and \DsspDssm fit fractions (in \%). Errors are statistical only.}
\label{tab:tab4}
\begin{center}
\vskip -0.2cm
\begin{tabular}{lccc}
\hline
\hline \noalign{\vskip2pt}
Resonance & & Fraction & \cr 
 & \DspDsm &  \DsspDsm & \DsspDssm  \cr
\hline \noalign{\vskip2pt}
$P(m)$ &  11 $\pm$ 5  & 27 $\pm$ 5 & 71 $\pm$ 20 \cr
$\psi(4040)$ & 62 $\pm$ 21 & & \cr
$\psi(4160)$ & 23 $\pm$ 26 &   53 $\pm$ 8 & \cr
$\psi(4415)$ & 6 $\pm$ 11 & 4 $\pm$ 2 & 5 $\pm$ 12 \cr
$X(4260)$ &  0.5 $\pm$ 3.0 & 18 $\pm$ 24 & 11 $\pm$ 16 \cr
\hline \noalign{\vskip2pt}
Sum & 103$\pm$ 36 & 102 $\pm$ 26& 87 $\pm$ 28\cr
\hline
\hline
\end{tabular}
\end{center}
\end{table}
The \DspDsm cross section is dominated by the $\psi(4040)$ resonance, and the \DsspDsm cross section by the $\psi(4160)$ resonance. The \DsspDssm cross section shows little resonance production.
The fits to the \DspDsm, \DsspDsm, and \DsspDssm mass spectra include the $X(4260)$ resonance, which is allowed to
interfere with all the other terms. In all cases, the $X(4260)$ 
fraction is consistent with zero. We note that the weak enhancement around 4.26 \gevcc in the  \DspDsm 
mass spectrum is described
by the fit in terms of interference between the $\psi(4040)$ and $\psi(4160)$ resonances. 

\section{Limits on $X(4260)$}

The $X(4260)$ yields are used to 
compute the cross section times branching fraction, to be compared with a \babar \ measurement of the 
$J/\psi\pi^+ \pi^-$ final state \cite{babar_Y}. 
The fractions from the fits reported in Table~\ref{tab:tab4} are converted to yields which are divided 
by the mass dependent  $\epsilon^{\cal{B}}$ efficiency and the integrated luminosity.
Systematic errors due to the mass and the width of the $\psi(4040)$,
$\psi(4160)$, $\psi(4415)$, and $X(4260)$ resonances are evaluated 
by varying the masses and widths by their uncertainty in the fit.
The size of the background contributions is varied within the statistical error, and 
the meson radii in the Breit-Wigner terms~\cite{blatt} are varied between 0 and 2.5 ${\rm GeV}^{-1}$.
Statistical and systematic errors are added in quadrature. 
We obtain
\begin{equation}
\frac{\BR(X(4260)\to \DspDsm)}{\BR(X(4260)\to J/\psi \pi^+ \pi^-)} < 0.7,
\end{equation}
\noindent
\begin{equation}
\frac{\BR(X(4260)\to \DsspDsm)}{\BR(X(4260)\to J/\psi \pi^+ \pi^-)} < 44,
\end{equation}
\noindent
and
\begin{equation}
\frac{\BR(X(4260)\to \DsspDssm)}{\BR(X(4260)\to J/\psi \pi^+ \pi^-)} < 30,
\end{equation}
\noindent
at the 95\% confidence level.

\section{Total cross section and Conclusion}
\begin{figure}[!htb]
\begin{center}
\includegraphics[width=10cm]{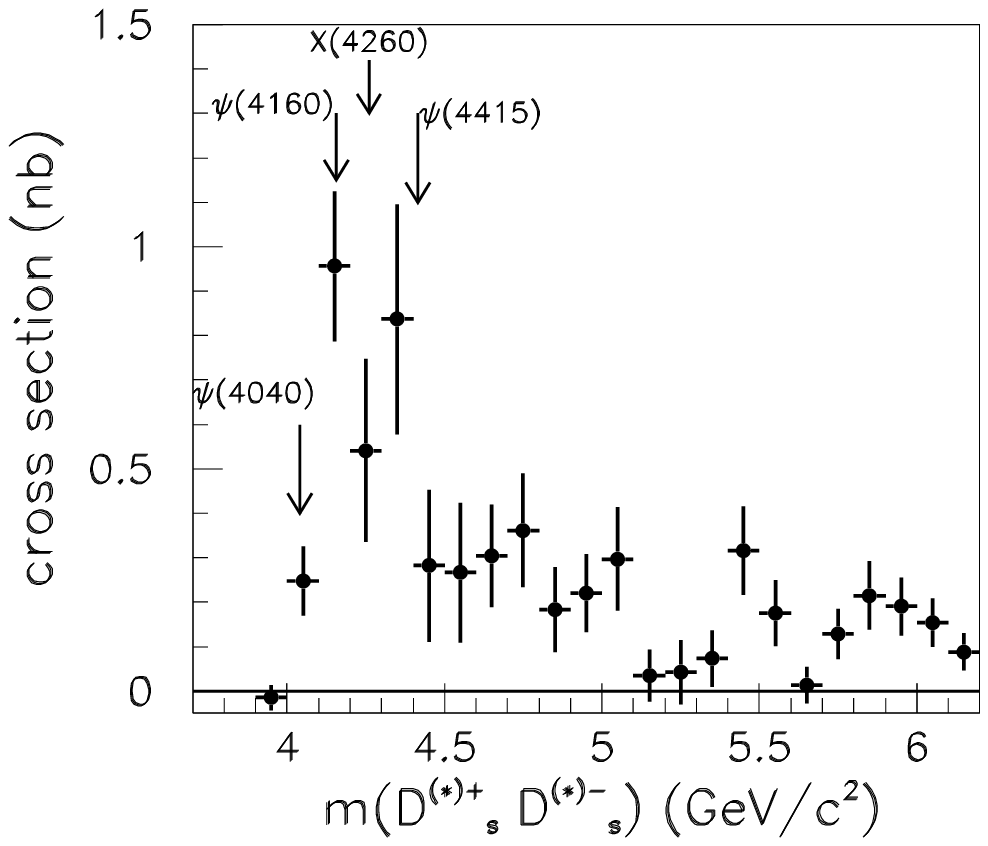}
\caption{Sum of $e^+ e^- \to  \DspDsm$, $e^+ e^- \to  \DsspDsm$, and $e^+ e^- \to  \DsspDssm$ cross sections.
Errors are statistical only.
The arrows indicate the positions of the different $\psi$ resonances and the $X(4260)$.} 
\label{fig:fig6}
\end{center}
\end{figure}

The sum of the $e^+ e^- \to  \DspDsm$, $e^+ e^- \to  \DsspDsm$, and 
$e^+ e^- \to  \DsspDssm$ cross sections 
is shown in Fig.~\ref{fig:fig6}; the arrows indicate the position of the 
different $\psi$ resonances and the $X(4260)$.
At the $X(4260)$ mass, there is a local minimum, similar to the measured
cross section for hadron production in $e^+e^-$ annihilation~\cite{pdg}.

In conclusion, we have studied the exclusive ISR production of the 
\DspDsm, \DsspDsm, and \DsspDssm final states.
The mass spectra show production of the $J^{PC}=1^{--}$ states,
$\psi(4040)$, $\psi(4160)$ and a weak indication for a smaller enhancement near 4.3 GeV.
From fits to the mass spectra for the three different final states we have
determined contributions by different $c \bar c$ resonances.

Upper limits on $X(4260)$ decays to these final states relative to $J/\psi \pip \pim$ are computed.
If the $X(4260)$ is a $1^{--}$ charmonium state, it should decay predominantly to open charm.
Within the present limited data sample size, no evidence is found for $X(4260)$ decays to \DspDsm, \DsspDsm, and \DsspDssm. If the X(4260) were a tetraquark state, it would decay
predominantly to \DspDsm~\cite{maiani}.

\section{Acknowledgements}
We are grateful for the 
extraordinary contributions of our \pep2\ colleagues in
achieving the excellent luminosity and machine conditions
that have made this work possible.
The success of this project also relies critically on the 
expertise and dedication of the computing organizations that 
support \babar.
The collaborating institutions wish to thank 
SLAC for its support and the kind hospitality extended to them. 
This work is supported by the
US Department of Energy
and National Science Foundation, the
Natural Sciences and Engineering Research Council (Canada),
the Commissariat \`a l'Energie Atomique and
Institut National de Physique Nucl\'eaire et de Physique des Particules
(France), the
Bundesministerium f\"ur Bildung und Forschung and
Deutsche Forschungsgemeinschaft
(Germany), the
Istituto Nazionale di Fisica Nucleare (Italy),
the Foundation for Fundamental Research on Matter (The Netherlands),
the Research Council of Norway, the
Ministry of Education and Science of the Russian Federation, 
Ministerio de Ciencia e Innovaci\'on (Spain), and the
Science and Technology Facilities Council (United Kingdom).
Individuals have received support from 
the Marie-Curie IEF program (European Union), the A. P. Sloan Foundation (USA) 
and the Binational Science Foundation (USA-Israel).


\begin{thebibliography}{99}
\bibitem{conj}
Charge conjugate states are implied throughout this work.
\end{thebibliography}
\end{document}